\documentclass[aps,prl,twocolumn,floatfix,,longbibliography]{revtex4-2}
\usepackage{amsmath,amssymb,graphicx,bm,color}
\def\veck{\mathbf k}
\def\vecq{\mathbf q}

\def\be{\begin{equation}}
\def\ee{\end{equation}}

\definecolor{darkred}{rgb}{0.6,0.0,0.0}

\begin{document}
\title{Anderson localization: A disorder-induced quantum bound state}

\author{V\'aclav  Jani\v{s}}  
\email{janis@fzu.cz}

\affiliation{Institute of Physics, The Czech Academy of Sciences, Na Slovance 1999/2, CZ-18200 Praha  8,  Czech Republic}
\date{\today}


\begin{abstract}
Electrons at the Fermi energy may lose their ability to propagate to long distances in certain random media. We use Green functions and solve parquet equations for the non-local electron-hole vertex in high spatial dimensions to describe the vanishing of diffusion in Anderson localization. It is caused by forming a quantum bound state between the diffusing particle and the hole left behind. Divergence in a new time scale proportional to the electrical polarizability signals the Anderson localization transition.  Consequently, the height of the peak of the dynamical conductivity at zero frequency, the static diffusion constant, is not pushed to zero at the localization transition but rather its width. Spatially localized quantum bound states in the localized phase cannot be described by the continuity and wave equations in the Hilbert space of Bloch waves.
\end{abstract}

\maketitle 
{\itshape Introduction.} 
Understanding charge diffusion in random media, such as metallic alloys, from the first quantum-mechanical principles has yet to be fully achieved. Since the seminal Anderson paper on the absence of diffusion in certain random lattices \cite{Anderson:1958aa}, a tremendous effort has been exerted to understand this concept \cite{Lee:1985aa,Evers:2008aa,IJMPB10}. Physical ideas lead Mott to suggest a minimal metallic conductivity at the mobility edge to Anderson localization \cite{Mott:1970aa}. Later, a minimal metallic conductivity and a first-order Anderson localization transition were challenged by scaling arguments of conductance with a continuous transition in dimensions $d>2$ \cite{Thouless:1974aa,Wegner:1976aa,Abrahams:1979aa}.   The continuous character of the Anderson localization transition was supported by transforming the original model near the mobility edge to a non-linear $\sigma$-model in $2 + \epsilon$ expansion \cite{Wegner:1979aa,Hikami:1981aa}. The nonexistence of the metallic phase in dimensions $d\le 2$ was corroborated by a self-consistent theory based on the perturbation expansion of the current relaxation kernel  using Green functions and Feynman diagrams \cite{Vollhardt:1980aa,Vollhardt:1980ab, Vollhardt:1992aa}. Anderson localization has also been intensively studied numerically on finite lattices \cite{Kramer:1993aa,Markos:2006aa}. The existence of a mobility edge separating localized from delocalized states of the $d>2$ Anderson model was proved rigorously \cite{Frohlich:1983aa}. Despite the numerous partial results, a complete microscopic theory of Anderson localization is missing. 

A suitable way to understand the microscopic origin of Anderson localization and to relate it to other critical phenomena is to use perturbation theory and Green functions. The first fully self-consistent perturbation theory for the electronic properties of random alloys is the coherent-potential approximation (CPA) \cite{Velicky:1968aa, Elliot:1974aa},  exact in infinite spatial dimensions, serving as a mean-field theory of single-electron properties \cite{Vlaming:1992aa,Janis:1992ab}. It misses, however, vertex corrections to the Drude conductivity and is unsuitable for Anderson localization \cite{Velicky:1969aa}. Backscattering and non-local correlations must be included to generate vertex corrections in the electrical conductivity \cite{Janis:2001aa, Janis:2003aa}. Self-consistent approximations for two-particle irreducible vertices must further be used when the Anderson localization transition should be disclosed in the renormalized perturbation theory  \cite{Janis:2001ab}. We used the parquet construction combining Bethe-Salpeter equations for nonlocal irreducible vertices in the electron-hole and electron-electron scattering channels, similar to Refs.~\cite{Vollhardt:1980aa,Vollhardt:1980ab, Vollhardt:1992aa}, who replaced the full two-particle self-consistency for the irreducible vertices by a self-consistent equation for the dynamical diffusion coefficient.   We found a critical point resembling the Anderson localization transition \cite{Janis:2005aa, Janis:2005ab}. The solution, however, did not obey the necessary Ward identity between the self-energy and the irreducible electron-hole vertex of Ref.~\cite{Vollhardt:1980ab}, guaranteeing the correct low-energy asymptotics of the diffusion pole in the density response function of the metallic phase.  
 
Generally, two fundamental properties must be met by reliable theories: causality of the two-particle vertex and Ward identity between the self-energy and the irreducible electron-hole vertex. They appear, however, incompatible in the perturbation expansion beyond the local CPA \cite{Janis:2004aa,Janis:2004ab}.  The diagrammatically derived vertex is causal from construction, but no self-energy can be found to obey the Ward identity with the causal non-local perturbative vertex. We later resolved this discrepancy by correcting the diagrammatic vertex in the appropriate two-particle subspace to fully comply with the Ward identity \cite{Janis:2016aa}. This rectification of the two-particle perturbation theory makes it possible to reinvestigate the connection between the earlier derived critical behavior in the Anderson model of disordered electrons and Anderson localization.
   
In this Letter, we derive a controllable conserving approximation for non-local two-particle irreducible vertices. It is generated by the leading terms of the self-consistent $1/d$ expansion and becomes asymptotically exact in high spatial dimensions. It obeys both mandatory consistency conditions of reliable theories: causality and Ward identity. It is qualitatively correct in all spatial dimensions with a transition to localized states in $d>2$,  no localized states in $ d=\infty$, and no delocalized states in $ d\le 2$. The Anderson localization transition is a non-equilibrium process with a crossover between tunneling and diffusive regimes of the metallic phase characterized by a new timescale. The localized states are found to be quantum-mechanical bound states confined in a finite volume delimited by a localization length diverging at the Anderson localization transition.  

{\itshape Model and Green functions}.
We use the Anderson model of an electron gas in a lattice random potential, the Hamiltonian of which, in terms of creation and annihilation operators of particles at site $\mathbf{R}_{i}$, is  
\begin{eqnarray}\label{eq:AD_hamiltonian}
\widehat{H} &=&\sum_{<ij>}t_{ij}\widehat{c}_{i}^{\dagger}
\widehat{c}_{j}+\sum_iV_i \widehat{c}_{i }^{\dagger } \widehat{c}_{i}\,.
\end{eqnarray}
The local potential $V_{i}$ is independently randomly distributed according to the same distribution. The two terms in the Hamiltonian do not commute, and the model is fully quantum-mechanical.

Assuming the ergodic hypothesis holds, we can express all physical quantities via averaged ones. The perturbation theory uses the non-random, translationally invariant propagator and expands in powers of the random potential. Each expansion term is then independently averaged. The appropriate quantities to be averaged are Green functions. In the case of the Fermi gas, they are only the one-particle Green function
\begin{subequations}
\begin{align}
 \label{eq:av_1PP}
  G_{ij}(z) &=
    \left\langle\left[z_1\widehat{1}-\widehat{H}\right]^{-1}_{ij}
    \right\rangle_{\rm av} 
\end{align}
and the two-particle one
\begin{align}
  \label{eq:av_2PP}
  G^{(2)}_{ij,kl}(z_1,z_2) &=
    \left\langle\left[z_1\widehat{1}-\widehat{H}\right]^{-1}_{ij}
    \left[z_2\widehat{1}-\widehat{H}\right]^{-1}_{kl}
    \right\rangle_{\rm av} \,,   
\end{align}
\end{subequations}
where $\langle \rangle_{av}$ denotes averaging over the random potential. The averaged Green functions are translationally invariant, just like the non-random ones. 

The impact of the random potential on the averaged one-particle Green function is contained in the self-energy $\Sigma_{\veck}(z)$  with wave vector $\veck$ and complex energy $z$. The particle's energy is conserved during the scatterings on the random potential for noninteracting systems. The two-particle Green function has two complex energies $z_{+}$ and $z_{-}$.  The physically relevant energies are the limits to real axis  $z_{\pm} \to E_{\pm}\pm i0^{+}$  producing retarded and advanced functions. We will use the following electron-hole Green function $\mathcal{G}^{RA}_{{\bf k}{\bf k}'}(E;\omega,{\bf q})$,  describing the propagation of a particle with energy $E_{+} =E + \omega/2 + i0^{+}$ and wave vector $\veck_{+}= \veck + \vecq/2 $ and a hole with energy $E_{-} = E  - \omega/2 - i0^{+}$ and wave vector $\veck_{-} = \veck - \vecq/2 $. The two-particle self-energy, the irreducible electron-hole vertex $L^{RA}_{{\bf k}{\bf k}'}(E;\omega,{\bf q})$ determines the averaged electron-hole propagator via a Bethe-Salpeter equation
\begin{multline}\label{eq:G2-momentum}
\mathcal{G}^{RA}_{{\bf k}{\bf k}'}(E;\omega,{\bf q}) = G_{\veck_{+}}(E_{+})G_{\veck_{-}}(E_{-}) \left[\delta(\veck - \veck^{\prime}) \phantom{\frac12}
\right. \\ \left.
 + \frac 1N \sum_{\mathbf{k}''} 
L^{RA}_{\mathbf{k}\mathbf{k}''}(E;\omega,\mathbf{q}) \mathcal{G}^{RA}_{\mathbf{k}''\mathbf{k}'}(E;\omega,\mathbf{q})\right]\,.
\end{multline}

The approximation is conserving if the difference of two self-energies $\Delta \Sigma_{\mathbf{k}}^{RA}(E;\omega,\mathbf{q})  = \Sigma_{\mathbf{k}_{+}}(E_{+}) - \Sigma_{\mathbf{k}_{-}}(E_{-})$ is related to the analogous difference of Green functions 
$\Delta G_{\mathbf{k}}^{RA}(E;\omega,\mathbf{q})  = G_{\mathbf{k}_{+}}(E_{+}) - G_{\mathbf{k}_{-}}(E_{-})$ via the electron-hole irreducible vertex \cite{Vollhardt:1980ab}
\begin{multline}\label{eq:WI-VW}
\Delta \Sigma_{\mathbf{k}}^{RA}(E;\omega,\mathbf{q})   
\\
= \frac 1{N}\sum_{\mathbf{k}'} L^{RA}_{\mathbf{k},\mathbf{k}'}(E, \omega;\mathbf{q}) 
\Delta G_{\mathbf{k}'}^{RA}(E;\omega,\mathbf{q}) \,.
\end{multline}
It guarantees that macroscopic conservation laws are obeyed.  

{\itshape Local mean-field solution}. 
The starting point for the perturbation theory for two-particle functions is the CPA, which is fully analytic, and all the irreducible functions are local and explicitly known. The self-energy $\Sigma(z)$ for complex energy $z$ is determined from the Soven equation
\begin{align} \label{eq:CPA_1}
 G(z)&=\left\langle\left[G^{-1}(z) + \Sigma(z) -
V_i\right]^{-1}\right\rangle_{av}
  \,,   
\end{align}
with the local Green function 
$ G(z) = \left\langle G\right\rangle\equiv N^{-1}\sum_{\veck} G_{\veck}(z)$.

The two-particle irreducible vertex in the complex plane of energies is 
\begin{subequations}
\begin{align}
\lambda(E;\omega) &= \frac {\Sigma(E_+) - \Sigma(E_-)} {G(E_+) - G(E_-)}  \,.
 \end{align}
It  complies with the Ward identity and determines the two-particle Green function with a restricted wave-vector dependence
 \begin{align}
  \mathcal{G}_{\veck\veck^{\prime}}^{RA}({E;\omega,\bf q}) &=   \frac{G_{\veck_{+}}(E_{+})G_{\veck_{-}}(E_{-})\delta(\veck - \veck^{\prime})}{1 -  \lambda(E;\omega)\chi^{RA}(E;\omega,\vecq)}   \,.
\end{align}
\end{subequations}
We denoted the electron-hole bubble $ \chi^{RA}(E;\omega,{\bf q}) = \left\langle G_{+}G_{-} \right\rangle \equiv N^{-1} \sum_{\veck } G_{\veck_{+}}(E_{+})G_{\veck_{-}}(E_{-} )$. Notice that the CPA vertex $\lambda \to \left\langle V_{i}^{2}\right\rangle_{av}$ in the weak-disorder limit.

{\itshape Two-particle self-consistency: High-dimensional critical behavior}. 
The CPA delivers a mean-field approximation and becomes exact for the local functions in $ d=\infty$. We must work with wave-vector-dependent self-energy and the two-particle irreducible vertices to go beyond it and reach the Anderson localization transition. It makes the approximations rather complex, but high spatial dimensions simplify the wave-vector dependence of the two-particle functions.  

We proposed a parquet approximation for the irreducible electron-hole vertex beyond the CPA. We found that a single, wave-vector-independent vertex $\Lambda_{0}(E,\omega)$ with the wave-vector dependent two-particle bubbles generates the leading $1/d$ contributions to the non-local two-particle irreducible vertices. The parquet equations solved in this high-dimensional approximation lead in systems with time-reversal symmetry to the following expression for the electron-hole irreducible vertex  \cite{Janis:2004aa,Janis:2004ab}
\begin{equation}\label{eq:parquet-momentum}
{\Lambda}_{\mathbf{k}\mathbf{k}'}(\mathbf{q}) = {\Lambda}
(\mathbf{q}) = \lambda + {\Lambda}_0 \frac {{\Lambda}_0 \bar{\chi}
(\vecq)} {1 - \Lambda_{0}\bar{\chi}( \vecq)}\,.
\end{equation}
We skipped the conserving energy variables to simplify the notation so that $\lambda^{RA}(E;\omega) \to \lambda$,  $\chi^{RA}(E;\omega,\vecq) \to \chi(\vecq)$ etc.  The bubble $\bar{\chi}(\vecq) = {\chi}(\vecq)  - \left\langle {\chi}(\vecq)\right\rangle_{\vecq}$ is a convolution of non-local Green funcions, $\left\langle F(\vecq)(\vecq)\right\rangle_{\vecq} = N^{-1}\sum_{\vecq}F(\vecq)$. 

The two-particle self-consistency is reached by a condition ${\Lambda}_0 = N^{-1}\sum_{\mathbf{q}}{\Lambda}(\mathbf{q})\equiv \left\langle \Lambda(\vecq)\right\rangle_{\vecq}$. The right-hand side of Eq.~\eqref{eq:parquet-momentum} contains a pole in the limit $\omega\to 0$ and $q\to 0$. The pole is integrable in dimensions $d>2$ and its frequency derivative in $d>4$, needed in the self-consistent solution and the accurate critical asymptotics. The qualitative, mean-field-like behavior will be guaranteed if we replace vertex $\Lambda_{0}$ in the denominator of the second term on the right-hand side by the CPA results for $\lambda$ and $\bar{\chi}(\vecq)$.  In this simplification, we do not change the qualitative behavior of vertex $\Lambda_{0}$ in $d>2$. It is then determined from a quadratic equation
\begin{equation}\label{eq:parquet-high_dim}
{\Lambda}_0 = \lambda + a {\Lambda}_0^2
\end{equation}
with 
\begin{align}
a &= \lambda \left[\left\langle\chi(\vecq)^{2}\right\rangle_{\vecq} - \left\langle \chi(\vecq)\right\rangle_{\vecq}^{2}\right] \,.
\end{align}
Notice that $a\propto d^{-1}$. Its physical root in the metallic regime, $4a<1$, is
\begin{align}\label{eq:Lambda0}
\Lambda_{0} &= \frac{1}{2a}\left[1 - \sqrt{1 - 4a\lambda}\right]\,.
\end{align}

This solution leads to a bifurcation point at $4a\lambda=1$ where the real vertex $\Lambda_{0}= \Lambda(E,\omega)$ splits at $\omega=0$ into two complex conjugate roots. We show that this is a hallmark of the Anderson localization transition. The critical value of the input CPA irreducible vertex at $\omega =0$ is
\be\label{eq:lambda-critical}
\lambda_{c}(E;0) = \frac 1{2\sqrt{\left\langle\chi(E;0,\vecq)^{2}\right\rangle_{\vecq} - \left\langle \chi(E;0,\vecq)\right\rangle_{\vecq}^{2}}} \,.
\ee
We see that  $\lambda_{c}\to\infty$ when $d\to\infty$.  
For the disorder strength $\lambda>\lambda_{c}$ the irreducible vertex becomes complex at zero transfer energy $\omega\to 0$ \cite{Janis:2005aa}
\begin{align}\label{eq:Lambda0I}
\Lambda_{0} 
&= \frac{1}{2a}\left[1 + i\,\mathrm{sign}(\omega) \sqrt{4a\lambda - 1}\right]\,.
\end{align}

{\itshape Conserving analytic approximation}. 
We solved the parquet equations for the irreducible electron-hole vertex $\Lambda_{\veck\veck^{\prime}}(\vecq) = \Lambda (\veck + \veck^{\prime} + \vecq)$  in high dimensional lattices with electron-hole symmetry. This vertex is causal but does not obey the desired Ward identity to make the approximation conserving. To make the integral Bethe-Salpeter equation for the corresponding conserving vertex analytically tractable and maintain its needed wave-vector-dependent critical behavior, we approximate the irreducible vertex in the following way  
\begin{align}\label{eq:lambdaq-approx}
\Lambda(\vecq) &= \lambda + \bar{\Lambda}\delta(\vecq) \,,
\end{align}
where $\bar{\Lambda} = \Lambda_{0} - \lambda>0$. It is again consistent with the leading $d^{-1}$ generic contribution to the irreducible two-particle vertex. The neglected contributions lead to wave-vector convolutions being of $1/d^{2}$ order or higher. 

According to Ref.~\cite{Janis:2016aa}, the conserving irreducible vertex constructed from the perturbative one, $\Lambda(\veck + \veck^{\prime} + \vecq)$, is in this approximation
\begin{multline}\label{eq:Lambda-conserving}
{L}_{\veck\veck^{\prime}}(\vecq) =  \lambda + \bar{\Lambda} \delta(\veck + \veck^{\prime} + \vecq) 
\\
- \frac 1{\left\langle \Delta G^{2}(\vecq)\right\rangle} 
\left[R_{\veck}(\vecq)\Delta G_{\veck^{\prime}}(\vecq)  + \Delta G_{\veck}(\vecq)R_{\veck^{\prime}}(\vecq) \phantom{\frac12}
\right.\\ \left.
 - \frac{\left\langle R(\vecq)\Delta G(\vecq)\right\rangle}{\left\langle \Delta G^{2}(\vecq)\right\rangle} \Delta G_{\veck}(\vecq)\Delta G_{\veck^{\prime}}(\vecq)\right] \,,
\end{multline}
 where the angular brackets denote normalized summation over the fermionic wave vectors $\left\langle F(\vecq)\right\rangle = N^{-1}\sum_{\veck}F_{\veck}(\vecq)$. 
The term restoring the Ward identity in this approximation  is
\begin{align}
R_{\veck}(\vecq) &= \lambda\left\langle \Delta G(\vecq) \right\rangle + \bar{\Lambda}\Delta G_{\veck}(\vecq) - \Delta\Sigma_{\veck}(\vecq)\,.
\end{align}

The two-particle approach has two-particle irreducible vertices as the fundamental entity for calculating all other quantities. In our parquet approximation they are $\lambda^{RA}(E,\omega)$ and $\bar{\Lambda}^{RA}(E,\omega)$. They are used to determine the imaginary part of the self-energy by using the Ward identity, Eq.~\eqref{eq:WI-VW} for  $\omega=0$ and $q=0$. We obtain  
\begin{align}\label{eq:ImSigmak}
\Im\Sigma_{\veck}(\omega_{+}) &= \frac{\lambda  \left\langle \Im G(\omega_{+})\right\rangle}{1 - \Re\bar{\Lambda}_{0} |G_{\veck}(\omega_{+})|^{2}}
\end{align}
where  $G_{\veck}(\omega_{+})= G^{R}_{\veck}(\omega) = [\omega + i0^{+} - \epsilon(\veck) - \Sigma_{\veck}(\omega + i0^{+})]^{-1}$. The real self-energy part is obtained from the Hilbert transform to keep the approximate Green functions analytic. It is a mean-field approximation where the self-energy depends on momenta only via the dispersion relation.

{\itshape Diffusion and Anderson localization}.
The full two-particle vertex $\mathcal{K}$ describing the effect of disorder on the two-particle propagation is obtained from the averaged two-particle Green function $\mathcal{G}_{\veck\veck^{\prime}}  = G_{\veck_{+}}G_{\veck_{-}}\left[\delta_{\veck,\veck^{\prime}} +  \mathcal{K}_{\veck\veck^{\prime}} G_{\veck^{\prime}_{+}}G_{\veck^{\prime}_{-}}\right]$. The  low-energy asymptotics of the electron-hole correlation function in the metallic phase is 
\begin{multline}\label{eq:Phi-D0}
\Phi^{RA}(E,\omega;\vecq) = \frac 1{N^{2}}\sum_{\veck,\veck^{\prime }}\mathcal{G}^{RA}_{{\bf k}{\bf k}'}(E;\omega,{\bf q})\  \xrightarrow[\omega,q\to0]{}
 \\ 
= \frac{2\pi n}{-i\omega + A(\omega)\omega^{2} + \bar{D}(\omega,\vecq)q^{2}}
\\
 = \frac{2\pi n(\omega)}{-i\omega + D(\omega,\vecq)q^{2} }\,,
\end{multline}
where $n= - \pi^{-1}\int_{-\infty}^{\infty}d\epsilon\rho(\epsilon)\Im\Sigma^{R}(\epsilon)/|E - \epsilon - \Sigma^{R}(\epsilon)|^{2}$ is the density of particles at energy $E$ and $\rho(\epsilon)$ is the bare density of states. The diffusion constant is the homogeneous, static limit of the general diffusion function, $D=\lim_{\omega,q\to 0} D(\omega,\vecq)$. The linear frequency term of the denominator does not depend on disorder due to the Ward identity. It is the new timescale $A = \lim_{\omega\to 0} A(E;\omega)>0$ at $\omega^{2}$ that is disorder-dependent and drives the system towards the Anderson localization transition. The canonical form of the diffusion pole, second equality, must use a complex frequency-dependent particle density $n(\omega) = n/(1 + iA\omega)$ and a dynamical diffusion function $D(\omega) \equiv D(\omega,\mathbf{0}) = D/(1 + iA\omega)$. The imaginary part of the dynamical particle density is the portion of particles not participating in the diffusion and not contributing to the conductivity. 
It is the dynamical conductivity $\sigma(\omega) = e^{2}n D(\omega)$ that contains the critical scale $A$ with a divergence in the frequency derivative $d\Im \sigma(\omega)/d\omega = - e^{2}AD n$ at $\omega=0$. The Drude peak of the dynamical conductivity vanishes at the Anderson localization transition not by pushing its height $\sigma_{0}$ to zero but by decreasing its width to zero.
 
Appendix contains the derivation of the conserving vertex from which we determine the new parameter $A$. Using the solution for vertex $\Lambda_{0}$ from Eq.~\eqref {eq:Lambda0} we obtain in the leading order near the critical point ($4a\lambda =1$) 
\begin{multline}\label{eq:A-D}
A =  \frac{\Im\dot{\bar{\Lambda}}\left\langle \Im G^{3}\right\rangle}{\lambda\left\langle \Im G\right\rangle\left\langle \Im G^{2}\right\rangle}
=\frac{\left\langle \Im G^{3}\right\rangle}{\left\langle \Im G\right\rangle\left\langle \Im G^{2} \right\rangle}
\\
\times
\left\langle \left\langle G_{\veck_{+}}G_{\veck_{-}} \Im\Delta G_{\veck}(\vecq)\right\rangle\left\langle G_{\veck_{+}}G_{\veck_{-}} - \left|\left\langle G\right\rangle\right|^{2}\right\rangle\right\rangle_{q}
\\
\times \frac{1 - 2 a\lambda - \sqrt{1 - 4a\lambda}}{2a^{2}\sqrt{1 - 4a\lambda}} \,.
\end{multline}
We can see that this coefficient diverges at the critical point when the input CPA irreducible vertex $\lambda_{c}(E;0)$ from Eq.~\eqref{eq:lambda-critical} reaches its critical value. The frequency dependence of $A_{c}(E,\omega)$  at the critical point is of order $|\omega|^{-1}$. It means that the denominator of the electron-hole correlation function is $\left[-i\omega + \alpha |\omega| + Dq^{2}\right]$. The coefficient $\alpha \in (0,\infty)$ depends on how we reach the critical point in the plane $[\lambda,\omega]$. Whereby $\alpha \propto |\omega|$ for $\lambda<\lambda_{c}$, $\alpha = \alpha_{c}$ for $\lambda = \lambda_{c}$, and $\alpha \propto |\omega|^{-1}$ for $\lambda>\lambda_{c}$. It means no pole in the electron-hole correlation function exists in the localized phase $\lambda>\lambda_{c}$.  

The static diffusion coefficient, $\omega=0$, in the leading order of this approximation reads
\begin{align}\label{eq:DiifusionConstant}
D &= - \left.\lambda\partial_{\vecq}^{2}\left\langle \gamma(\vecq) \Delta G(\vecq) \right\rangle\right|_{q=0} 
\nonumber \\
&=  \lambda^{2}\left|\left\langle \Im G\right\rangle\right| \left\langle \frac{1 + \Re\bar{\Lambda}_{0}\left|G\right|^{2}}{\left(1 - \Re\bar{\Lambda}_{0}\left|G\right|^{2}\right)^{3}}\left|\nabla G\right|^{2}\right\rangle\,, 
\end{align}
where $\gamma_{\veck}(\vecq) = 1/\left(\lambda\left\langle \Delta G\right\rangle + \Delta\epsilon -  \bar{\Lambda}\Delta G_{\veck}(\vecq)\right)$ and we jused Eq.~\eqref{eq:ImSigmak}, see the Appendix. We denoted the differential operator $\nabla \equiv \nabla_{\veck}$.

Parameter $A$ introduces a new time scale, the relevance of which increases when approaching the critical point. The low-energy limit of the electron-hole correlation function corresponds to the long-time regime. It is described in the space-time representation by a diffusion equation. However, we must correct the standard diffusion equation and add a term with the second derivative in time to include the new time scale $A$. We then must use the following long-time asymptotic behavior of the particle density or particle probability distribution
\be\label{eq:Diffusion-A}
\frac{\partial}{\partial t} n(t,\mathbf{x})  = \left[A\frac{\partial^{2}}{\partial t^{2}} + D\nabla_{\mathbf{x}}^{2}\right] n(t,\mathbf{x}) \,. 
\ee
This is an extended diffusion equation for the long-time limit of particle propagation in the metallic phase. It combines classical diffusion with quantum tunneling. When the first-order time derivative dominates, we are in the diffusive regime of a metallic phase. On the other hand, when the second-order time derivative overtakes control, the moving particle is confined in a spatially restricted area and can reach infinite distance only via tunneling through an effective potential barrier.    

The explicit solution of Eq.~\eqref{eq:Diffusion-A}  in time can be obtained by using contour integration in the inverse Fourier transform from frequencies and wave vectors.  The general forward propagator, $t>0$, is    
\begin{equation}\label{eq:Diff-AM}
\widetilde{n}(t,\vecq) =  \widetilde{n}(0,\vecq)
\frac{\exp\left(-\displaystyle{\frac{t}{2A}}\left(\sqrt{4 ADq^{2} + 1} - 1\right)\right)}{\sqrt{4 ADq^{2} + 1}} \,.
\end{equation}

The new disorder-dependent time scale $A$ affects not only the low-frequency asymptotics of $D(\omega,\vecq)$ as discussed below Eq.~\eqref{eq:Phi-D0}.  A dimensionless parameter $ADq^{2}$ dominates the small-momentum asymptotics of the static diffusion function $D(\vecq) \equiv D(0,\vecq)$. In the weak-disorder limit, $ADq^{2}\ll 1$, the diffusion coefficient is lowered to $D(\vecq) \to D\left(1 - ADq^{2}\right)$. In the strong-disorder limit, $ADq^{2}\gg 1$, it turns to be $D(\vecq) \to D/\sqrt{ADq^{2}}$.  The scale $A$ becomes experimentally relevant when $A\approx L^{2}/D\approx n_{F}L^{2}/\left\langle (\nabla\epsilon)^{2} \Im G^{2}\right\rangle$. Similarly, it becomes relevant on the time scale $t\ll A$ on which deviations from the standard diffusion can be observed.     

The metallic, diffusive regime vanishes beyond the Anderson localization transition with $A=\infty$ at $\lambda_{c}$ from Eq.~\eqref{eq:lambda-critical}. The Fourier transform of the electron-hole correlation function contains no diffusion pole in the localized phase. Its low-energy asymptotic form below the Anderson localization transition is
\be\label{eq:Diffusion-L}
\Phi^{RA}(E,\omega;\vecq) = \frac{2\pi n_{F}}{-i\omega + D\left(\xi^{-2}  + q^{2}\right)}
\ee
We introduced a localization length $\xi$ defined as   
\begin{equation}\label{eq:xi-lambda}
\xi^{-2}
= - \frac{\left(4a\lambda - 1 \right)\left\langle \Im G^{3}\right\rangle}{4a^{2}\lambda D\left\langle \Im G\right\rangle^{2}} \,,
 \end{equation}
as derived in the Appendix. It means that there is no diffusion pole in the localized phase. The inverse localization length determines the distance to the pole and the probability $p(x) \propto e^{-x/\xi}$ of finding the particle at distance $x$ from the origin after infinite waiting time.  There is a threshold voltage $U_{0} = h^{2}\xi^{-2}/2m|e|$ above which the electric current can flow in the localized phase and the system is metallic, characterized by a modified Ohm's law $I(U) = e^{2}nD\theta(U - U_{0})(U - U_{0})$. No current at any electric field in the localized phase exists in theories with vanishing static conductivity $\sigma_{0} = e^{2} nD =0$. The electron in the localized phase forms a bound state with the hole left behind at the origin and the binding energy of this state is $E_{0} = U_{0}/|e|$.    

{\itshape Conclusions}.
We solved the parquet equations, self-consistently interconnecting multiple scattering in the electron-electron (Cooperon) and electron-hole (diffuson) singular channels, for the non-local two-particle irreducible vertices of the disordered Anderson model in a high-dimensional $d^{-1}$ expansion. We constructed a two-particle mean-field theory where higher-order $O(d^{-2})$ momentum convolutions were neglected. We calculated the electron-hole vertex and the electron-hole correlation function. They contain a diffusion pole in the metallic phase, the low-energy limit of which was determined to the second order in frequency. The disorder-dependent quadratic term is critical at the Anderson localization transition with $A\omega^{2} \xrightarrow[\omega\to 0]{} D/\xi^{2}$ in the localized phase and the localization length $\xi\le\infty$. Coefficient $A$ is proportional to the real part of the static electrical polarizability $\Re\alpha(\omega) = -\lim_{\omega\to0}\Im\sigma(\omega)/\omega = \sigma_{0}A/(1 + A^{2}\omega^{2})$. It causes the width of the dynamical conductivity to shrink to zero and the frequency derivative and second momentum derivative of the diffusion function $D(\omega,\vecq)$ at $\omega=0$ and $q=0$ to diverge at the Anderson localization transition. The existing approaches miss this parameter and are left with the static diffusion coefficient $D$ as the only disorder-dependent quantity. Its vanishing is then the only way to define the Anderson localization transition. The diffusion pole in these approaches survives in the localized phase, which contradicts its integrability enforced by the two-particle self-consistency of the parquet equations \cite{Janis:2009aa}.  Consequently, the static electrical conductivity calculated from the Kubo formula does not reach zero at the transition from the metallic phase.   

We conclude that the Anderson localization transition in the conserving mean-field theory is a non-equilibrium dynamic process in which a new time scale  $A$ beyond the linear response theory diverges at the transition. It must be added to the diffusion equation as a second-order time derivative. Anderson localization is a quantum-mechanical bound state outside the Bloch waves' Hilbert space. There is no diffusion in the localized phase, but a strong electric field above a threshold value determined by the binding energy of the localized state can enforce charge diffusion and turn the localized phase metallic.

\section*{Acknowledgment}
 I thank Dieter Vollhardt for the valuable and inspiring discussions. 
 

\begin{thebibliography}{30}%
\makeatletter
\providecommand \@ifxundefined [1]{%
 \@ifx{#1\undefined}
}%
\providecommand \@ifnum [1]{%
 \ifnum #1\expandafter \@firstoftwo
 \else \expandafter \@secondoftwo
 \fi
}%
\providecommand \@ifx [1]{%
 \ifx #1\expandafter \@firstoftwo
 \else \expandafter \@secondoftwo
 \fi
}%
\providecommand \natexlab [1]{#1}%
\providecommand \enquote  [1]{``#1''}%
\providecommand \bibnamefont  [1]{#1}%
\providecommand \bibfnamefont [1]{#1}%
\providecommand \citenamefont [1]{#1}%
\providecommand \href@noop [0]{\@secondoftwo}%
\providecommand \href [0]{\begingroup \@sanitize@url \@href}%
\providecommand \@href[1]{\@@startlink{#1}\@@href}%
\providecommand \@@href[1]{\endgroup#1\@@endlink}%
\providecommand \@sanitize@url [0]{\catcode `\\12\catcode `\$12\catcode
  `\&12\catcode `\#12\catcode `\^12\catcode `\_12\catcode `\%12\relax}%
\providecommand \@@startlink[1]{}%
\providecommand \@@endlink[0]{}%
\providecommand \url  [0]{\begingroup\@sanitize@url \@url }%
\providecommand \@url [1]{\endgroup\@href {#1}{\urlprefix }}%
\providecommand \urlprefix  [0]{URL }%
\providecommand \Eprint [0]{\href }%
\providecommand \doibase [0]{https://doi.org/}%
\providecommand \selectlanguage [0]{\@gobble}%
\providecommand \bibinfo  [0]{\@secondoftwo}%
\providecommand \bibfield  [0]{\@secondoftwo}%
\providecommand \translation [1]{[#1]}%
\providecommand \BibitemOpen [0]{}%
\providecommand \bibitemStop [0]{}%
\providecommand \bibitemNoStop [0]{.\EOS\space}%
\providecommand \EOS [0]{\spacefactor3000\relax}%
\providecommand \BibitemShut  [1]{\csname bibitem#1\endcsname}%
\let\auto@bib@innerbib\@empty
\bibitem [{\citenamefont {Anderson}(1958)}]{Anderson:1958aa}%
  \BibitemOpen
  \bibfield  {author} {\bibinfo {author} {\bibfnamefont {P.~W.}\ \bibnamefont
  {Anderson}},\ }\bibfield  {title} {\bibinfo {title} {Absence of diffusion in
  certain random lattices},\ }\href {https://doi.org/10.1103/PhysRev.109.1492}
  {\bibfield  {journal} {\bibinfo  {journal} {Phys. Rev.}\ }\textbf {\bibinfo
  {volume} {109}},\ \bibinfo {pages} {1492} (\bibinfo {year}
  {1958})}\BibitemShut {NoStop}%
\bibitem [{\citenamefont {Lee}\ and\ \citenamefont
  {Ramakrishnan}(1985)}]{Lee:1985aa}%
  \BibitemOpen
  \bibfield  {author} {\bibinfo {author} {\bibfnamefont {P.~A.}\ \bibnamefont
  {Lee}}\ and\ \bibinfo {author} {\bibfnamefont {T.~V.}\ \bibnamefont
  {Ramakrishnan}},\ }\bibfield  {title} {\bibinfo {title} {Disordered
  electronic systems},\ }\href {https://doi.org/10.1103/RevModPhys.57.287}
  {\bibfield  {journal} {\bibinfo  {journal} {Reviews of Modern Physics}\
  }\textbf {\bibinfo {volume} {57}},\ \bibinfo {pages} {287} (\bibinfo {year}
  {1985})}\BibitemShut {NoStop}%
\bibitem [{\citenamefont {Evers}\ and\ \citenamefont
  {Mirlin}(2008)}]{Evers:2008aa}%
  \BibitemOpen
  \bibfield  {author} {\bibinfo {author} {\bibfnamefont {F.}~\bibnamefont
  {Evers}}\ and\ \bibinfo {author} {\bibfnamefont {A.}~\bibnamefont {Mirlin}},\
  }\bibfield  {title} {\bibinfo {title} {Anderson transitions},\ }\href
  {https://doi.org/10.1103/RevModPhys.80.1355} {\bibfield  {journal} {\bibinfo
  {journal} {Reviews of Modern Physics}\ }\textbf {\bibinfo {volume} {80}},\
  \bibinfo {pages} {1355} (\bibinfo {year} {2008})}\BibitemShut {NoStop}%
\bibitem [{\citenamefont {Abrahams}(2010)}]{IJMPB10}%
  \BibitemOpen
  \bibinfo {editor} {\bibfnamefont {E.}~\bibnamefont {Abrahams}},\ ed.,\
  \href@noop {} {\emph {\bibinfo {title} {50 Years of Anderson Localization}}}\
  (\bibinfo  {publisher} {World Scientific Publishing, Singapore},\ \bibinfo
  {year} {2010})\BibitemShut {NoStop}%
\bibitem [{\citenamefont {Mott}(1970)}]{Mott:1970aa}%
  \BibitemOpen
  \bibfield  {author} {\bibinfo {author} {\bibfnamefont {N.~F.}\ \bibnamefont
  {Mott}},\ }\bibfield  {title} {\bibinfo {title} {Conduction in
  non-crystalline systems: Iv. anderson localization in a disordered lattice},\
  }\href {https://doi.org/10.1080/14786437008228147} {\bibfield  {journal}
  {\bibinfo  {journal} {Philosophical Magazine}\ }\textbf {\bibinfo {volume}
  {22}},\ \bibinfo {pages} {7} (\bibinfo {year} {1970})}\BibitemShut {NoStop}%
\bibitem [{\citenamefont {Thouless}(1974)}]{Thouless:1974aa}%
  \BibitemOpen
  \bibfield  {author} {\bibinfo {author} {\bibfnamefont {D.~J.}\ \bibnamefont
  {Thouless}},\ }\bibfield  {title} {\bibinfo {title} {Electrons in disordered
  systems and the theory of localization},\ }\href
  {https://doi.org/10.1016/0370-1573(74)90029-5} {\bibfield  {journal}
  {\bibinfo  {journal} {Physics Reports}\ }\textbf {\bibinfo {volume} {13}},\
  \bibinfo {pages} {93} (\bibinfo {year} {1974})}\BibitemShut {NoStop}%
\bibitem [{\citenamefont {Wegner}(1976)}]{Wegner:1976aa}%
  \BibitemOpen
  \bibfield  {author} {\bibinfo {author} {\bibfnamefont {F.~J.}\ \bibnamefont
  {Wegner}},\ }\bibfield  {title} {\bibinfo {title} {Electrons in disordered
  systems. scaling near the mobility edge},\ }\href
  {https://doi.org/10.1007/BF01315248} {\bibfield  {journal} {\bibinfo
  {journal} {Zeitschrift f{\"u}r Physik B Condensed Matter}\ }\textbf {\bibinfo
  {volume} {25}},\ \bibinfo {pages} {327} (\bibinfo {year} {1976})}\BibitemShut
  {NoStop}%
\bibitem [{\citenamefont {Abrahams}\ \emph {et~al.}(1979)\citenamefont
  {Abrahams}, \citenamefont {Anderson}, \citenamefont {Licciardello},\ and\
  \citenamefont {Ramakrishnan}}]{Abrahams:1979aa}%
  \BibitemOpen
  \bibfield  {author} {\bibinfo {author} {\bibfnamefont {E.}~\bibnamefont
  {Abrahams}}, \bibinfo {author} {\bibfnamefont {P.~W.}\ \bibnamefont
  {Anderson}}, \bibinfo {author} {\bibfnamefont {D.~C.}\ \bibnamefont
  {Licciardello}},\ and\ \bibinfo {author} {\bibfnamefont {T.~V.}\ \bibnamefont
  {Ramakrishnan}},\ }\bibfield  {title} {\bibinfo {title} {Scaling theory of
  localization: Absence of quantum diffusion in two dimensions},\ }\href
  {https://doi.org/10.1103/PhysRevLett.42.673} {\bibfield  {journal} {\bibinfo
  {journal} {Physical Review Letters}\ }\textbf {\bibinfo {volume} {42}},\
  \bibinfo {pages} {673} (\bibinfo {year} {1979})}\BibitemShut {NoStop}%
\bibitem [{\citenamefont {Wegner}(1979)}]{Wegner:1979aa}%
  \BibitemOpen
  \bibfield  {author} {\bibinfo {author} {\bibfnamefont {F.}~\bibnamefont
  {Wegner}},\ }\bibfield  {title} {\bibinfo {title} {The mobility edge problem:
  Continuous symmetry and a conjecture},\ }\href
  {https://doi.org/10.1007/BF01319839} {\bibfield  {journal} {\bibinfo
  {journal} {Zeitschrift f{\"u}r Physik B Condensed Matter}\ }\textbf {\bibinfo
  {volume} {35}},\ \bibinfo {pages} {207} (\bibinfo {year} {1979})}\BibitemShut
  {NoStop}%
\bibitem [{\citenamefont {Hikami}(1981)}]{Hikami:1981aa}%
  \BibitemOpen
  \bibfield  {author} {\bibinfo {author} {\bibfnamefont {S.}~\bibnamefont
  {Hikami}},\ }\bibfield  {title} {\bibinfo {title} {Anderson localization in a
  nonlinear-sigma-model reprentation},\ }\href
  {https://doi.org/10.1103/PhysRevB.24.2671} {\bibfield  {journal} {\bibinfo
  {journal} {Physical Review B}\ }\textbf {\bibinfo {volume} {24}},\ \bibinfo
  {pages} {2671} (\bibinfo {year} {1981})}\BibitemShut {NoStop}%
\bibitem [{\citenamefont {Vollhardt}\ and\ \citenamefont
  {W\"olfle}(1980{\natexlab{a}})}]{Vollhardt:1980aa}%
  \BibitemOpen
  \bibfield  {author} {\bibinfo {author} {\bibfnamefont {D.}~\bibnamefont
  {Vollhardt}}\ and\ \bibinfo {author} {\bibfnamefont {P.}~\bibnamefont
  {W\"olfle}},\ }\bibfield  {title} {\bibinfo {title} {Anderson localization in
  $d\le 2$ dimensions: A self-consistent diagrammatic theory},\ }\href
  {https://doi.org/10.1103/PhysRevLett.45.842} {\bibfield  {journal} {\bibinfo
  {journal} {Phys. Rev. Lett.}\ }\textbf {\bibinfo {volume} {45}},\ \bibinfo
  {pages} {842} (\bibinfo {year} {1980}{\natexlab{a}})}\BibitemShut {NoStop}%
\bibitem [{\citenamefont {Vollhardt}\ and\ \citenamefont
  {W\"olfle}(1980{\natexlab{b}})}]{Vollhardt:1980ab}%
  \BibitemOpen
  \bibfield  {author} {\bibinfo {author} {\bibfnamefont {D.}~\bibnamefont
  {Vollhardt}}\ and\ \bibinfo {author} {\bibfnamefont {P.}~\bibnamefont
  {W\"olfle}},\ }\bibfield  {title} {\bibinfo {title} {Diagrammatic,
  self-consistent treatment of the {Anderson} localization problem in $d\le 2$
  dimensions},\ }\href {https://doi.org/10.1103/PhysRevB.22.4666} {\bibfield
  {journal} {\bibinfo  {journal} {Phys. Rev. B}\ }\textbf {\bibinfo {volume}
  {22}},\ \bibinfo {pages} {4666} (\bibinfo {year}
  {1980}{\natexlab{b}})}\BibitemShut {NoStop}%
\bibitem [{\citenamefont {Vollhardt}\ and\ \citenamefont
  {W\"olfle}(1992)}]{Vollhardt:1992aa}%
  \BibitemOpen
  \bibfield  {author} {\bibinfo {author} {\bibfnamefont {D.}~\bibnamefont
  {Vollhardt}}\ and\ \bibinfo {author} {\bibfnamefont {P.}~\bibnamefont
  {W\"olfle}},\ }\bibfield  {title} {\bibinfo {title} {Self-consistent theory
  of anderson localization},\ }in\ \href@noop {} {\emph {\bibinfo {booktitle}
  {Electronic Phase Transitions}}},\ \bibinfo {editor} {edited by\ \bibinfo
  {editor} {\bibfnamefont {W.}~\bibnamefont {Hanke}}\ and\ \bibinfo {editor}
  {\bibfnamefont {{\relax Yu}.~V.}\ \bibnamefont {Kopaev}}}\ (\bibinfo
  {publisher} {Elsevier Science Publishers B. V., Amsterodam},\ \bibinfo {year}
  {1992})\ Chap.~\bibinfo {chapter} {1}, pp.\ \bibinfo {pages}
  {1--78}\BibitemShut {NoStop}%
\bibitem [{\citenamefont {Kramer}\ and\ \citenamefont
  {MacKinnon}(1993)}]{Kramer:1993aa}%
  \BibitemOpen
  \bibfield  {author} {\bibinfo {author} {\bibfnamefont {B.}~\bibnamefont
  {Kramer}}\ and\ \bibinfo {author} {\bibfnamefont {A.}~\bibnamefont
  {MacKinnon}},\ }\bibfield  {title} {\bibinfo {title} {Localization: theory
  and experiment},\ }\href {https://doi.org/10.1088/0034-4885/56/12/001}
  {\bibfield  {journal} {\bibinfo  {journal} {Rep. Prog. Phys.}\ }\textbf
  {\bibinfo {volume} {56}},\ \bibinfo {pages} {1469} (\bibinfo {year}
  {1993})}\BibitemShut {NoStop}%
\bibitem [{\citenamefont {{Marko\v s}}(2006)}]{Markos:2006aa}%
  \BibitemOpen
  \bibfield  {author} {\bibinfo {author} {\bibfnamefont {P.}~\bibnamefont
  {{Marko\v s}}},\ }\bibfield  {title} {\bibinfo {title} {Numerical analysis of
  the anderson localization},\ }\href@noop {} {\bibfield  {journal} {\bibinfo
  {journal} {Acta Phys. Slovaca}\ }\textbf {\bibinfo {volume} {56}},\ \bibinfo
  {pages} {561} (\bibinfo {year} {2006})}\BibitemShut {NoStop}%
\bibitem [{\citenamefont {Fr{\"o}hlich}\ and\ \citenamefont
  {Spencer}(1983)}]{Frohlich:1983aa}%
  \BibitemOpen
  \bibfield  {author} {\bibinfo {author} {\bibfnamefont {J.}~\bibnamefont
  {Fr{\"o}hlich}}\ and\ \bibinfo {author} {\bibfnamefont {T.}~\bibnamefont
  {Spencer}},\ }\bibfield  {title} {\bibinfo {title} {Absence of diffusion in
  the anderson tight abdence of diffusion in the anderson tight binding model
  for large disorder or low energy},\ }\href@noop {} {\bibfield  {journal}
  {\bibinfo  {journal} {Communications in Mathematical Physics}\ }\textbf
  {\bibinfo {volume} {88}},\ \bibinfo {pages} {151} (\bibinfo {year}
  {1983})}\BibitemShut {NoStop}%
\bibitem [{\citenamefont {Velick{\'y}}\ \emph {et~al.}(1968)\citenamefont
  {Velick{\'y}}, \citenamefont {Kirkpatrick},\ and\ \citenamefont
  {Ehrenreich}}]{Velicky:1968aa}%
  \BibitemOpen
  \bibfield  {author} {\bibinfo {author} {\bibfnamefont {B.}~\bibnamefont
  {Velick{\'y}}}, \bibinfo {author} {\bibfnamefont {S.}~\bibnamefont
  {Kirkpatrick}},\ and\ \bibinfo {author} {\bibfnamefont {H.}~\bibnamefont
  {Ehrenreich}},\ }\bibfield  {title} {\bibinfo {title} {Single-site
  approximations in the electronic theory of simple binary alloys},\ }\href
  {https://doi.org/10.1103/PhysRev.175.747} {\bibfield  {journal} {\bibinfo
  {journal} {Physical Review}\ }\textbf {\bibinfo {volume} {175}},\ \bibinfo
  {pages} {747} (\bibinfo {year} {1968})}\BibitemShut {NoStop}%
\bibitem [{\citenamefont {Elliott}\ \emph {et~al.}(1974)\citenamefont
  {Elliott}, \citenamefont {Krumhansl},\ and\ \citenamefont
  {Leath}}]{Elliot:1974aa}%
  \BibitemOpen
  \bibfield  {author} {\bibinfo {author} {\bibfnamefont {R.~J.}\ \bibnamefont
  {Elliott}}, \bibinfo {author} {\bibfnamefont {J.~A.}\ \bibnamefont
  {Krumhansl}},\ and\ \bibinfo {author} {\bibfnamefont {P.~L.}\ \bibnamefont
  {Leath}},\ }\bibfield  {title} {\bibinfo {title} {The theory and properties
  of randomly disordered crystals and related physical systems},\ }\href
  {https://doi.org/10.1103/RevModPhys.46.465} {\bibfield  {journal} {\bibinfo
  {journal} {Rev. Mod. Phys.}\ }\textbf {\bibinfo {volume} {46}},\ \bibinfo
  {pages} {465} (\bibinfo {year} {1974})}\BibitemShut {NoStop}%
\bibitem [{\citenamefont {Vlaming}\ and\ \citenamefont
  {Vollhardt}(1992)}]{Vlaming:1992aa}%
  \BibitemOpen
  \bibfield  {author} {\bibinfo {author} {\bibfnamefont {R.}~\bibnamefont
  {Vlaming}}\ and\ \bibinfo {author} {\bibfnamefont {D.}~\bibnamefont
  {Vollhardt}},\ }\bibfield  {title} {\bibinfo {title} {Controlled mean-field
  theory for disordered electronic systems: Single-particle properties},\
  }\href {https://doi.org/10.1103/PhysRevB.45.4637} {\bibfield  {journal}
  {\bibinfo  {journal} {Physical Review B}\ }\textbf {\bibinfo {volume} {45}},\
  \bibinfo {pages} {4637} (\bibinfo {year} {1992})}\BibitemShut {NoStop}%
\bibitem [{\citenamefont {Jani{\v s}}\ and\ \citenamefont
  {Vollhardt}(1992)}]{Janis:1992ab}%
  \BibitemOpen
  \bibfield  {author} {\bibinfo {author} {\bibfnamefont {V.}~\bibnamefont
  {Jani{\v s}}}\ and\ \bibinfo {author} {\bibfnamefont {D.}~\bibnamefont
  {Vollhardt}},\ }\bibfield  {title} {\bibinfo {title} {Coupling of quantum
  degrees of freedom in strongly interacting disordered electron systems},\
  }\href {https://doi.org/10.1103/PhysRevB.46.15712} {\bibfield  {journal}
  {\bibinfo  {journal} {Physical Review B}\ }\textbf {\bibinfo {volume} {46}},\
  \bibinfo {pages} {15712} (\bibinfo {year} {1992})}\BibitemShut {NoStop}%
\bibitem [{\citenamefont {Velick{\'y}}(1969)}]{Velicky:1969aa}%
  \BibitemOpen
  \bibfield  {author} {\bibinfo {author} {\bibfnamefont {B.}~\bibnamefont
  {Velick{\'y}}},\ }\bibfield  {title} {\bibinfo {title} {Theory of electronic
  transport in disordered binary alloys: Coherent-potential approximation},\
  }\href {https://doi.org/10.1103/PhysRev.184.614} {\bibfield  {journal}
  {\bibinfo  {journal} {Physical Review}\ }\textbf {\bibinfo {volume} {184}},\
  \bibinfo {pages} {614} (\bibinfo {year} {1969})}\BibitemShut {NoStop}%
\bibitem [{\citenamefont {Jani{\v s}}\ and\ \citenamefont
  {Vollhardt}(2001)}]{Janis:2001aa}%
  \BibitemOpen
  \bibfield  {author} {\bibinfo {author} {\bibfnamefont {V.}~\bibnamefont
  {Jani{\v s}}}\ and\ \bibinfo {author} {\bibfnamefont {D.}~\bibnamefont
  {Vollhardt}},\ }\bibfield  {title} {\bibinfo {title} {Conductivity of
  disordered electrons: Mean-field approximation containing vertex
  corrections},\ }\href {https://doi.org/10.1103/PhysRevB.63.125112} {\bibfield
   {journal} {\bibinfo  {journal} {Physical Review B}\ }\textbf {\bibinfo
  {volume} {63}},\ \bibinfo {pages} {125112} (\bibinfo {year}
  {2001})}\BibitemShut {NoStop}%
\bibitem [{\citenamefont {Jani{\v s}}\ \emph {et~al.}(2003)\citenamefont
  {Jani{\v s}}, \citenamefont {Koloren{\v c}},\ and\ \citenamefont {{\v S}pi{\v
  c}ka}}]{Janis:2003aa}%
  \BibitemOpen
  \bibfield  {author} {\bibinfo {author} {\bibfnamefont {V.}~\bibnamefont
  {Jani{\v s}}}, \bibinfo {author} {\bibfnamefont {J.}~\bibnamefont {Koloren{\v
  c}}},\ and\ \bibinfo {author} {\bibfnamefont {V.}~\bibnamefont {{\v S}pi{\v
  c}ka}},\ }\bibfield  {title} {\bibinfo {title} {Density and current response
  functions in strongly disordered electron systems: diffusion, electrical
  conductivity and einstein relation},\ }\href@noop {} {\bibfield  {journal}
  {\bibinfo  {journal} {European Physical Journal B}\ }\textbf {\bibinfo
  {volume} {35}},\ \bibinfo {pages} {77} (\bibinfo {year} {2003})}\BibitemShut
  {NoStop}%
\bibitem [{\citenamefont {Jani{\v s}}(2001)}]{Janis:2001ab}%
  \BibitemOpen
  \bibfield  {author} {\bibinfo {author} {\bibfnamefont {V.}~\bibnamefont
  {Jani{\v s}}},\ }\bibfield  {title} {\bibinfo {title} {Parquet approach to
  nonlocal vertex functions and electrical conductivity of disordered
  electrons},\ }\href {https://doi.org/10.1103/PhysRevB.64.115115} {\bibfield
  {journal} {\bibinfo  {journal} {Physical Review B}\ }\textbf {\bibinfo
  {volume} {64}},\ \bibinfo {pages} {115115} (\bibinfo {year}
  {2001})}\BibitemShut {NoStop}%
\bibitem [{\citenamefont {Jani{\v s}}\ and\ \citenamefont {Koloren{\v
  c}}(2005{\natexlab{a}})}]{Janis:2005aa}%
  \BibitemOpen
  \bibfield  {author} {\bibinfo {author} {\bibfnamefont {V.}~\bibnamefont
  {Jani{\v s}}}\ and\ \bibinfo {author} {\bibfnamefont {J.}~\bibnamefont
  {Koloren{\v c}}},\ }\bibfield  {title} {\bibinfo {title} {Mean-field theory
  of anderson localization: Asymptotic solution in high spatial dimensions},\
  }\href {https://doi.org/10.1103/PhysRevB.71.033103} {\bibfield  {journal}
  {\bibinfo  {journal} {Physical Review B}\ }\textbf {\bibinfo {volume} {71}},\
  \bibinfo {pages} {033103} (\bibinfo {year} {2005}{\natexlab{a}})}\BibitemShut
  {NoStop}%
\bibitem [{\citenamefont {Jani{\v s}}\ and\ \citenamefont {Koloren{\v
  c}}(2005{\natexlab{b}})}]{Janis:2005ab}%
  \BibitemOpen
  \bibfield  {author} {\bibinfo {author} {\bibfnamefont {V.}~\bibnamefont
  {Jani{\v s}}}\ and\ \bibinfo {author} {\bibfnamefont {J.}~\bibnamefont
  {Koloren{\v c}}},\ }\bibfield  {title} {\bibinfo {title} {Mean-field theories
  for disordered electrons: Diffusion pole and anderson localization},\ }\href
  {https://doi.org/10.1103/PhysRevB.71.245106} {\bibfield  {journal} {\bibinfo
  {journal} {Physical Review B}\ }\textbf {\bibinfo {volume} {71}},\ \bibinfo
  {pages} {245106} (\bibinfo {year} {2005}{\natexlab{b}})}\BibitemShut
  {NoStop}%
\bibitem [{\citenamefont {Jani{\v s}}\ and\ \citenamefont {Koloren{\v
  c}}(2004{\natexlab{a}})}]{Janis:2004aa}%
  \BibitemOpen
  \bibfield  {author} {\bibinfo {author} {\bibfnamefont {V.}~\bibnamefont
  {Jani{\v s}}}\ and\ \bibinfo {author} {\bibfnamefont {J.}~\bibnamefont
  {Koloren{\v c}}},\ }\bibfield  {title} {\bibinfo {title} {Conservation laws
  in disordered electron systems: Thermodynamic limit and configurational
  averaging},\ }\href@noop {} {\bibfield  {journal} {\bibinfo  {journal}
  {Physica Status Solidi (b)}\ }\textbf {\bibinfo {volume} {241}},\ \bibinfo
  {pages} {2032} (\bibinfo {year} {2004}{\natexlab{a}})}\BibitemShut {NoStop}%
\bibitem [{\citenamefont {Jani{\v s}}\ and\ \citenamefont {Koloren{\v
  c}}(2004{\natexlab{b}})}]{Janis:2004ab}%
  \BibitemOpen
  \bibfield  {author} {\bibinfo {author} {\bibfnamefont {V.}~\bibnamefont
  {Jani{\v s}}}\ and\ \bibinfo {author} {\bibfnamefont {J.}~\bibnamefont
  {Koloren{\v c}}},\ }\bibfield  {title} {\bibinfo {title} {Causality versus
  ward identity in disordered electron systems},\ }\href@noop {} {\bibfield
  {journal} {\bibinfo  {journal} {Modern Physics Letters B}\ }\textbf {\bibinfo
  {volume} {18}},\ \bibinfo {pages} {1051} (\bibinfo {year}
  {2004}{\natexlab{b}})}\BibitemShut {NoStop}%
\bibitem [{\citenamefont {Jani{\v s}}\ and\ \citenamefont {Koloren{\v
  c}}(2016)}]{Janis:2016aa}%
  \BibitemOpen
  \bibfield  {author} {\bibinfo {author} {\bibfnamefont {V.}~\bibnamefont
  {Jani{\v s}}}\ and\ \bibinfo {author} {\bibfnamefont {J.}~\bibnamefont
  {Koloren{\v c}}},\ }\bibfield  {title} {\bibinfo {title} {Conserving
  approximations for response functions of the fermi gas in a random
  potential},\ }\href@noop {} {\bibfield  {journal} {\bibinfo  {journal}
  {European Physical Journal B}\ }\textbf {\bibinfo {volume} {89}},\ \bibinfo
  {pages} {1434} (\bibinfo {year} {2016})}\BibitemShut {NoStop}%
\bibitem [{\citenamefont {Jani{\v s}}(2009)}]{Janis:2009aa}%
  \BibitemOpen
  \bibfield  {author} {\bibinfo {author} {\bibfnamefont {V.}~\bibnamefont
  {Jani{\v s}}},\ }\bibfield  {title} {\bibinfo {title} {Integrability of the
  diffusion pole in the diagrammatic description of noninteracting electrons in
  a random potential},\ }\href {https://doi.org/10.1088/0953-8984/21/48/485501}
  {\bibfield  {journal} {\bibinfo  {journal} {Journal of Physics: Condensed
  Matter}\ }\textbf {\bibinfo {volume} {21}},\ \bibinfo {pages} {485501}
  (\bibinfo {year} {2009})}\BibitemShut {NoStop}%
\end{thebibliography}

%

\onecolumngrid\newpage

\section{Appendix}

\twocolumngrid
{\itshape Perturbative vertex}.
The standard perturbation theory is applied to the one-particle self-energy, from which all other quantities are derived via exact mathematical relations. When we need a two-particle self-consistency, however, the corresponding irreducible vertex becomes the central quantity of the perturbation theory. The self-energy and the one-particle propagators must then be reconstructed from the two-particle irreducible vertex. The self-energy and the irreducible vertex are connected via a Ward identity in conserving approximations. 
However, the dynamic approximations cannot generically guarantee that the Ward identity is fully obeyed in individual microscopic scattering processes since the two-particle vertex contains more information than the self-energy. Only the reduced Ward identity, Eq.~\eqref{eq:ImSigmak}, can be used in the approximations with a known two-particle vertex.  Using the electron-hole irreducible vertex from Eq.~\eqref{eq:lambdaq-approx}, the full perturbative vertex is determined from the Bethe-Salpeter equation 
%
%
\begin{multline}\label{eq:2Pvertex-pertrbative}
\Gamma_{\mathbf{k}\mathbf{k}'}(E;\omega,\mathbf{q}) =
  {\Lambda}_{\mathbf{k}\mathbf{k}'}(E;\omega,\mathbf{q})
  + \frac
  1N\sum_{\mathbf{k}''}
  {\Lambda}_{\mathbf{k}\mathbf{k}''}(E;\omega,\mathbf{q}) %
  \\ \times
  {G}_{\mathbf{k}_{+}''}(E_{+}) {G}_{\mathbf{k}_{-}''}(E_{-})
  \Gamma_{\mathbf{k}''\mathbf{k}'}(E;\omega,\mathbf{q}) \,,
\end{multline}
where $\veck_{\pm} = \veck \pm \vecq/2$, $E_{\pm} = E \pm \omega/2 \pm i0^{+}$.
Its homogeneous version with  $q=0$ in approximation from Eq.~\eqref{eq:lambdaq-approx}  is
\begin{widetext}
\begin{multline}
\Gamma_{\veck\veck^{\prime}}(E;\omega,\mathbf{0}) = \frac{1}{\left[1 - \bar{\Lambda}(E;\omega) G_{\veck}(E_{+})G_{\veck}(E_{-})\right]\left[1 - \bar{\Lambda}(E;\omega) G_{\veck^{\prime}}(E_{+})G_{\veck^{\prime}}(E_{-})\right]} 
 \\ 
\times \left\{\bar{\Lambda}(E;\omega)\delta_{\veck,\veck^{\prime}}\left[1 - \bar{\Lambda}(E;\omega) G(E_{+})G(E_{-})\right] + \frac{\lambda(E;\omega)}{1 - \lambda(E;\omega) \displaystyle{\left\langle\frac{G(E_{+})G(E_{-})}{\left[1 - \bar{\Lambda}(E;\omega) G(E_{+})G(E_{-})\right]} \right\rangle}} \right\} \,, 
\end{multline}
\end{widetext}
where the angular brackets denote averaging over fermionic wave vectors. 
It is easy to show by using Eq.~\eqref{eq:ImSigmak} that this dynamic vertex is divergent for $\omega=0$ since
\begin{align}
\lambda\left\langle\frac{| G|^{2}}{1 - \bar{\Lambda}| G|^{2}}\right\rangle &= \frac{\left\langle |G|^{2} \Im \Sigma\right\rangle }{\left\langle \Im G\right\rangle} =1 \,.
\end{align}
 The dynamic vertex from the perturbation theory contains the singularity of the diffusion pole. Still, it does not reproduce the low-energy asymptotics $\omega\to 0$ as needed due to the complete Ward identity, Eq.~\eqref{eq:WI-VW}. The physical, conserving vertex obeys the Bethe-Salpeter equation with the irreducible vertex $L_{\veck\veck^{\prime}}(E;\omega,\vecq)$.

{\itshape  Conserving vertex}. 
We use the following representations to determine the low-energy asymptotics of the diffusion pole in the conserving vertex
\begin{subequations}
\begin{multline}
G^{RA}_{\veck}(E;\omega,\vecq) \equiv G^{R}_{\veck_{+}}(E_{+})G^{A}_{\veck_{-}}(E_{-}) 
\\
= \frac{\Delta G_{\veck}(\vecq)}{\lambda \left\langle \Delta G(\vecq) \right\rangle + \bar{\Lambda}\Delta G_{\veck}(\vecq) - R_{\veck}(\vecq) + \Delta\epsilon_{\veck}(\vecq) - \omega}  
\end{multline}
and
\begin{align}\label{eq:gamma}
\frac{G^{RA}}{1 - \bar{\Lambda} G^{RA}} 
&= \frac{\Delta G}{\lambda \left\langle \Delta G\right\rangle - R + \Delta\epsilon - \omega} \equiv \gamma \Delta G \,.
\end{align}
\end{subequations}
We denoted $\Delta\epsilon_{\veck}(\vecq) = \epsilon(\veck_{+}) - \epsilon(\veck_{-}) $. Suppressing the energy variables, the Bethe-Salpeter equation for the full vertex is  
\begin{widetext}
\begin{multline}\label{eq:K-def}
\mathcal{K}_{\veck\veck^{\prime}}(\vecq) = \frac 1{1 - \bar{\Lambda}G^{RA}_{\veck}(\vecq)}\left\{ \lambda + \bar{\Lambda} \delta(\veck + \veck^{\prime} + \vecq) 
- \frac 1{\left\langle \Delta G^{2}(\vecq)\right\rangle} 
\left[R_{\veck}(\vecq)\Delta G_{\veck^{\prime}}(\vecq) + \Delta G_{\veck}(\vecq)R_{\veck^{\prime}}(\vecq)  \phantom{\frac12}
\right.\right. \\ \left.\left.
-\ \frac{\left\langle R(\vecq)\Delta G(\vecq)\right\rangle}{\left\langle \Delta G^{2}(\vecq)\right\rangle} \Delta G_{\veck}(\vecq)\Delta G_{\veck^{\prime}}(\vecq)\right] + \lambda\left\langle G^{RA}(\vecq)\mathcal{K}_{\veck^{\prime}}(\vecq)\right\rangle
- \frac 1{\left\langle \Delta G^{2}(\vecq)\right\rangle} 
\left[R_{\veck}(\vecq) \left\langle \Delta G(\vecq)G^{RA}(\vecq)\mathcal{K}_{\veck^{\prime}}(\vecq)\right\rangle \phantom{\frac12}
\right.\right. \\ \left.\left.
+\ \Delta G_{\veck}(\vecq)\left(\left\langle R(\vecq)G^{RA}(\vecq)\mathcal{K}_{\veck^{\prime}}(\vecq)\right\rangle - \frac{\left\langle R(\vecq)\Delta G(\vecq)\right\rangle}{\left\langle \Delta G^{2}(\vecq)\right\rangle}\left\langle \Delta G(\vecq)G^{RA}(\vecq)\mathcal{K}_{\veck^{\prime}}(\vecq)\right\rangle \,,
\right) 
\right]
\right\}
\end{multline}
\end{widetext}
where
$
\left\langle F\mathcal{K}_{\veck^{\prime}}\right\rangle = N^{-1}\sum_{\veck}F_{\veck}\mathcal{K}_{\veck\veck^{\prime}}$. The explicit expression for vertex $\mathcal{K}_{\veck\veck^{\prime}}(\vecq)$ is obtained from solving an algebraic matrix equation for its three contractions
 $\left\langle G^{RA}(\vecq)\mathcal{K}_{\veck^{\prime}}(\vecq)\right\rangle$, $\left\langle R(\vecq)G^{RA}(\vecq)\mathcal{K}_{\veck^{\prime}}(\vecq)\right\rangle$, and $\left\langle \Delta G(\vecq)G^{RA}(\vecq)\mathcal{K}_{\veck^{\prime}}(\vecq)\right\rangle$. Vertex $\mathcal{K}_{\veck\veck^{\prime}}(\vecq)$ contains the same singularity as the dynamic one $\Gamma_{\veck\veck^{\prime}}(\vecq)$ for $\omega=0$. It moreover restores the exact linear asymptotics $\omega \to 0$.

{\itshape Low-energy asymptotics}.
We obtain the diffusion pole's asymptotic form from the determinant of the matrix determining vertex $\mathcal{K}$ from Eq.~\eqref{eq:K-def}.  We set $\vecq =\mathbf{0}$ and use the symmetry $\epsilon(\veck) = \epsilon(-\veck)$. We expand the determinant to the second order in frequency.  After resolving $\left\langle G^{RA}(\mathbf{0})\mathcal{K}_{\veck^{\prime}}(\mathbf{0})\right\rangle$, $\left\langle R(\mathbf{0})G^{RA}(\mathbf{0})\mathcal{K}_{\veck^{\prime}}(\vecq)\right\rangle$, and $\left\langle \Delta G(\mathbf{0})G^{RA}(\mathbf{0})\mathcal{K}_{\veck^{\prime}}(\mathbf{0})\right\rangle$ in Eq.~\eqref{eq:K-def} we obtain the asymptotic form of the denominator of the conserving vertex near the Anderson localization transition
\begin{widetext}
\begin{multline}
\mathcal{D}(\omega) \equiv \mathcal{D}(E;\omega,\mathbf{0}) \doteq 1 - \lambda\left\langle \gamma \Delta G\right\rangle 
+\ \frac{\omega}{\left\langle \Delta G^{2}\right\rangle} \left[\left(1 - \lambda\left\langle \gamma\Delta G\right\rangle \right)\left(2\left\langle  \dot{R}\gamma\Delta G^{2}\right\rangle - \frac{\left\langle \dot{R} \Delta G\right\rangle}{\left\langle\Delta G^{2}\right\rangle}\left\langle \gamma \Delta G^{3}\right\rangle\right) 
\right. \\ \left.
+\ \lambda\left\langle \gamma \Delta G^{2}\right\rangle\left(2\left\langle  \dot{R}\gamma\Delta G\right\rangle - \frac{\left\langle \dot{R} \Delta G\right\rangle}{\left\langle\Delta G^{2}\right\rangle}\left\langle \gamma \Delta G^{2}\right\rangle\right) \right]
+\   \frac{\omega^{2}}{\left\langle \Delta G^{2}\right\rangle^{2}}
\left\{
\left(1 - \lambda\left\langle \gamma \Delta G\right\rangle\right)\left[ \left\langle\dot{R} \gamma \Delta G^{2}\right\rangle^{2} - \left\langle\dot{R}^{2} \gamma \Delta G\right\rangle \left\langle\gamma \Delta G^{3}\right\rangle 
 \right]
 \right.\\ \left.
+\ \lambda\left\langle \dot{R}\gamma\Delta G\right\rangle\left[
 2 \left\langle \gamma\Delta G^{2}\right\rangle \left\langle \dot{R}\gamma\Delta G^{2}\right\rangle - \left\langle \gamma\Delta G^{3}\right\rangle \left\langle \dot{R} \gamma \Delta G\right\rangle
 \right]  - \lambda\left\langle \gamma\Delta G^{2}\right\rangle^{2}\left\langle \dot{R}^{2}\gamma\Delta G\right\rangle
 \right\} \,,
 \end{multline}
\end{widetext}
where we used $\dot{R} \doteq i\Im\dot{\bar{\Lambda}}\Delta G$  with $\Delta G = 2i\Im G^{R}$ at $\omega=0$.
We further expand the remaining frequency-dependent function 
\begin{align}\label{eq:gammaExpansion}
\gamma &\doteq \frac1{\lambda\left\langle \Delta G\right\rangle}\left[ 1 + \frac{1 + \dot{R}}{\lambda\left\langle\Delta G\right\rangle}\omega + \frac{\left(1 + \dot{R}\right)^{2}}{\lambda^{2}\left\langle\Delta G\right\rangle^{2}}\omega^{2}\right] \,.
\end{align}
All the expansion coefficients are now taken at $\omega = 0$. Assuming $\dot{R}\gg 1$,
the low-energy asymptotics of the denominator of the conserving vertex and the diffusion pole then is  
 \begin{multline}\label{eq:calDomega}
 \mathcal{D}(\omega) \doteq - \frac{\omega}{\lambda \left\langle \Delta G\right\rangle}    
 -  \frac{i\Im\dot{\bar{\Lambda}}\left\langle\Delta G^{3}\right\rangle\omega^{2}}{\lambda^{2}\left\langle\Delta G\right\rangle^{2}\left\langle\Delta G^{2}\right\rangle}\  
 \\
 =
   \frac{-1}{2\lambda \left\langle \Im G\right\rangle}\left[-i\omega + \frac{\Im\dot{\bar{\Lambda}}\left\langle \Im G^{3}\right\rangle\omega^{2}}{\lambda\left\langle \Im G\right\rangle\left\langle \Im G^{2}\right\rangle}\right] \,,
 \end{multline}
The linear term in frequency does not depend on the disorder strength, as demanded from the exact theory. The second-order term does not lose its dependence on disorder even if the Ward identity, Eq.~\eqref{eq:WI-VW}, is fully obeyed. It is proportional to function $R_{\veck}(E,0)$, containing the information from the electron-hole irreducible vertex that is not reflected in the self-energy.  Its non-zero value means the disorder drives the states toward localized bound states. Unless the second-order contribution diverges, the states are localized only temporarily. 

The corresponding low-energy asymptotics of the electron-hole correlation function in the metallic phase is
\begin{subequations}\label{eq:Phi-diffusive}
\begin{multline}\label{eq:Phi_{D}}
\Phi^{RA}(E,\omega;\vecq) = \frac 1{N^{2}}\sum_{\veck,\veck^{\prime }}\mathcal{G}^{RA}_{{\bf k}{\bf k}'}(E;\omega,{\bf q}) 
\\
= -\frac{1}{\lambda\mathcal{D}(E;\omega,\vecq)}
\xrightarrow[\omega,q\to 0]{} \frac{2\pi n}{-i\omega + A\omega^{2} + Dq^{2}}\,,
\end{multline}
where 
$n = -\left\langle \Im G\right\rangle/\pi$ is the renormalized density of particles.
The scale $A$ diverges at the Anderson localization transition, and the electron-hole correlation function in the localized phase is
\begin{align}
\Phi^{RA}(E,\omega;\vecq)  &\doteq \frac{2\pi n}{-i\omega + D\left(\xi^{-2} + q^{2}\right)} \,,
\end{align}
\end{subequations}
where $D>0$ is the asymptotic value of the static diffusion constant at the Anderson localization transition, Eq.   ~\eqref{eq:DiifusionConstant}. The inverse localization length  is obtained  from Eqs.~\eqref{eq:gammaExpansion} and~\eqref{eq:calDomega}  if  $\dot{R}_{\veck}(\mathbf{0})\omega \to - 2\Im\Lambda_{0}\Im G_{\veck}$ for $\omega= 0$. The result close to the transition point then is 
\begin{align}\label{eq:kappa-def}
\xi^{-2} =  \frac{\Im\Lambda_{0}^{2}\left\langle\left| \Im G^{3}\right|\right\rangle}{\lambda D\left\langle \Im G\right\rangle^{2}} \,\
\end{align}
assuming $\Im\bar{\Lambda}_{0}\ll \lambda$. Inserting the solution of Eq.~\eqref{eq:Lambda0I}  we obtain the result in Eq.~\eqref{eq:xi-lambda}.
\end{document}